\documentclass[aps,twocolumn,superscriptaddress,showpacs,prb]{revtex4}
\usepackage{graphicx}
\usepackage{amsmath}
\usepackage{nccmath}
\newcommand{\msr}{$\mu$SR}
\newcommand{\tio}{TiO$_2$}
\begin{document}



\title{Electronic structure of Mu-complex donor state in rutile TiO$_2$}


\author{K. Shimomura}
\affiliation{Muon Science Laboratory, Institute of Materials Structure Science, High Energy Accelerator Research Organization (KEK), Tsukuba, Ibaraki 305-0801, Japan}
\affiliation{School of Mathematical and Physical Science, The Graduate University for Advanced Studies, Tsukuba, Ibaraki 305-0801, Japan}
\author{R. Kadono}\thanks{Corresponding author: ryosuke.kadono@kek.jp}
\affiliation{Muon Science Laboratory, Institute of Materials Structure Science, High Energy Accelerator Research Organization (KEK), Tsukuba, Ibaraki 305-0801, Japan}
\affiliation{School of Mathematical and Physical Science, The Graduate University for Advanced Studies, Tsukuba, Ibaraki 305-0801, Japan}
\author{A. Koda}
\affiliation{Muon Science Laboratory, Institute of Materials Structure Science, High Energy Accelerator Research Organization (KEK), Tsukuba, Ibaraki 305-0801, Japan}
\affiliation{School of Mathematical and Physical Science, The Graduate University for Advanced Studies, Tsukuba, Ibaraki 305-0801, Japan}
\author{K. Nishiyama}
\affiliation{Muon Science Laboratory, Institute of Materials Structure Science, 
High Energy Accelerator Research Organization (KEK), Tsukuba, Ibaraki 305-0801, Japan}
\author{M. Mihara}
\affiliation{Department of Physics, Osaka University, Toyonaka, Osaka 560-0043, Japan}


\date{\today}

\begin{abstract}
The hyperfine structure of the interstitial muonium (Mu) in rutile (TiO$_2$, weakly $n$-type) has been identified by means of a muon spin rotation technique. The angle-resolved hyperfine parameters exhibit a tetragonal anisotropy within the $ab$ plane and axial anisotropy with respect to the $\langle 001\rangle$ ($\hat{c}$) axis. This strongly suggests that the Mu is bound to O (forming an OH bond) at an off-center site within a channel along the $\hat{c}$ axis, while the unpaired Mu electron is localized around the neighboring Ti site. The hyperfine parameters are quantitatively explained by a model that considers spin polarization of the unpaired electron at both the Ti and O sites, providing evidence for the formation of Mu as a Ti-O-Mu complex paramagnetic state.  The disappearance of the Mu signal above $\sim$10 K suggests that the energy necessary for the promotion of the unpaired electron to the conduction band by thermal activation is of the order of $10^1$ meV. These observations suggest that, while the electronic structure of Mu (and hence H) differs considerably from that of the conventional shallow level donor described by the effective mass model,  Mu supplies a loosely bound electron, and thus, serves as a donor in rutile.
\end{abstract}

\pacs{61.72.Hh, 71.55.Ht, 76.75.+i}

\maketitle




\section{Introduction}
Hydrogen (H) is a ubiquitous impurity in a wide variety of semiconductors, including elemental (e.g., Si) and binary compound (e.g., GaAs) materials, which comprise the primary basis for current industrial applications. In these covalent systems, H is known to be amphoteric, forming an acceptor or donor level in $n$- or $p$-type materials, respectively \cite{Patterson:88,Pankove:90,Myers:92,Chow:98}.  Meanwhile, a recent theoretical prediction that H could behave as an independent electron donor to induce $n$-type conductivity in ZnO \cite{Walle:00}, along with subsequent support from various experiments involving muon spin rotation (\msr) \cite{Cox:01,Shimomura:02}, electron paramagnetic resonance (EPR),  electron-nuclear double resonance (ENDOR) measurements \cite{Hofmann:02}, and additional conventional techniques \cite{Gil:99,Shi:04,Vilao:05}, have prompted the development of a generalized hypothesis of H as a shallow donor dopant in wide gap semiconductors \cite{Kilic:02,Walle:03,Peacock:03}.

As in the case of ZnO, \msr\ studies have made a significant contribution to this subject by providing spectroscopic information on muonium (Mu; an analog of a neutral H atom where the proton is replaced by a positive muon) observed in a wide variety of semiconductors. While the dynamical aspects (e.g., diffusion) of Mu and H may differ considerably, because of the relatively low mass of Mu ($m_\mu \simeq m_p/9$, with $m_p$ being the proton mass), the local electronic structure of Mu is virtually equivalent to that of H if a small correction to account for the difference in the reduced mass ($\sim$0.5\%) is made. The recent discovery of novel Mu states having extremely small hyperfine parameters ($\sim$10$^{-4}$ times smaller than that of Mu in vacuum) and low ionization energy ($\sim$10$^1$ meV) in several compound semiconductors, including CdS \cite{Gil:99}, ZnSe \cite{Vilao:05}, InN \cite{Davis:03}, and GaN \cite{Shimomura:04}, also supports the hypothesis that Mu (and hence H) can act as a donor in these compounds. 

The generalized hypothesis also predicts shallow donor H in various metal oxides, most notably in rutile \tio, where the origin of the unintentional $n$-type conductivity exhibited by the as-grown crystal is the focus of intense research. While the electrical and optical properties of \tio\ are subject to both intrinsic and extrinsic defects, this material has a rich variety of potential applications in both electronic and opto-electronic devises, and detailed understanding of the H behavior is crucially important to the material functionality in this context.

\begin{figure}[t]
\includegraphics[width=0.95\linewidth]{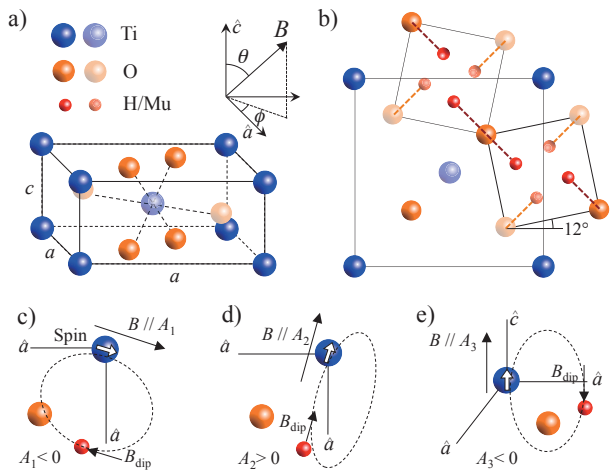}
\caption{(Color online) (a) Schematic illustration of rutile \tio\ crystal structure, showing a unit cell with $a=b=0.4593$ nm and $c=0.2959$ nm. (b) Atomic position of interstitial hydrogen (H), where H forms OH bonds (0.109 nm) at four inequivalent H sites.
(c)--(e) Signs of hyperfine parameters ($A_i$) expected for the magnetic dipolar field ($B_{\rm dip}$) generated by an electron spin localized on Ti atom under an external field ($\vec{B}$). (see text).}
\label{fig1}
\end{figure}

As shown in Fig.~\ref{fig1}(a), rutile has a tetragonal structure with Ti$^{4+}$ ions occupying the body-center position, where each Ti ion is coordinated by six ligand oxygen atoms (O$^{2-}$) comprising a slightly distorted octahedron. The TiO$_6$ octahedra are linked by corner-sharing to form O chains parallel to the $\hat{c}$ axis. Regions of low electron density exist between the chains, forming the so-called ``$c$ channel".  Earlier studies have shown that interstitial H resides in the $c$ channel, leading to the formation of an OH bond (with a length of 0.109 nm) perpendicular to the $\hat{c}$ axis \cite{Anderson:73}.  The open channels cause anisotropy in the diffusion process, and may allow fast diffusion of smaller ions parallel to the tetragonal axis.  It has recently been reported that the H diffusion can be strongly enhanced via irradiation of the crystal by infrared light that matches the OH bond stretch mode \cite{Spahr:10}.

Here, we report on the electronic structure of Mu in single-crystalline rutile. Although the existence of a Mu state with an extremely small hyperfine (HF) parameter has been reported elsewhere \cite{Shimomura:05,Cox:06}, little is known about the local electronic structure.  In this study, we show that the HF parameters exhibit strong anisotropy within the $ab$ plane as well as along the $\hat{c}$ axis. The angular dependence of the HF parameters is perfectly in line with that expected for Mu occupying the sites that are common to the interstitial H and forming OMu bonds.  More interestingly, the relative signs of these parameters indicate that the HF interaction is predominantly determined by the magnetic dipolar fields generated by the unpaired Mu electron located away from the muon site.  This is qualitatively consistent with a recent report on the H-related paramagnetic center, in which the HF interaction is attributed to the Ti$^{3+}$[OH]$^-$ complex state with the $d$ electron being located on the Ti ion nearest to the OH base \cite{Brant:11}.  Meanwhile, a detailed comparison of the Mu and H HF structures reveals a distinct difference between the complex centers,  where a contribution from the residual spin polarization at the nearest neighboring O site (antiparallel to that of the Ti ion) is suggested in the Mu case.  Through consideration of the temperature ($T$) dependence of the Mu yield, we also show that the unpaired electron bound to the Mu complex state requires small energy for promotion to the conduction band.  Based on these observations, we argue that interstitial Mu/H is one of the primary origins of unintentional $n$-type conductivity in rutile \tio. This is despite its complex electronic structure, which differs significantly from that for the conventional effective-mass-like donor.

\section{Experiment}
A conventional \msr\ experiment was conducted at the former Booster Meson Facility in Tsukuba (at KEK) and at J-PARC MUSE.  A 4-MeV muon beam that was almost 100\% spin-polarized parallel to the beam direction was implanted into single-crystalline rutile wafers [30 mm $\times$ 30 mm, 1.0-mm thick, grown using the Verneuil process], which were ``as received" from a local vendor (Furuuchi Chemical Co.).  The definitions of the azimuthal and polar angles ($\phi$, $\theta$, respectively) for the external magnetic field ($\vec{B}$) are shown in Fig.~\ref{fig1}(a). For the $\phi$-dependence measurements, a wafer having a normal axis orientation of $\langle001\rangle$ ($\parallel \hat{c}$) was used, where the initial muon spin polarization ($\vec{P}_\mu$) was parallel to $\hat{c}$, and the $\hat{a}$ axis was rotated contrary to the $\vec{B}$ direction within the $ab$ plane (i.e., $\theta=90^\circ$). Meanwhile, the $\theta$-dependence was measured by tilting the $\vec{B}$ direction towards the $\hat{c}$ axis (except for $\theta=0$, for which another $\langle100\rangle$ crystal was used), where the angle was tuned by an additional field parallel to the $\hat{c}$ axis.  In both cases, the $\vec{B}$-parallel component of the HF parameter was observed in terms of the frequency shift of the satellite signals. 

\section{Results}
\begin{figure}
\includegraphics[width=0.95\linewidth]{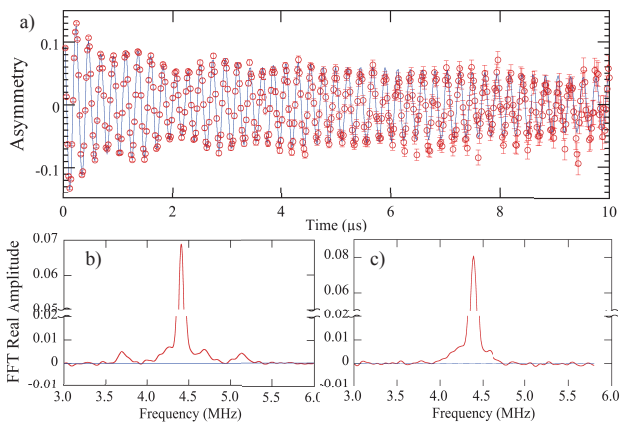}
\caption{(Color online) (a) Example of \msr\ time spectrum observed in \tio\ at 5 K under an applied magnetic field of 32.5 mT, where the field ($\vec{B}$) is rotated from $\langle100\rangle$ axis ($\hat{a}$) by $\phi=30^\circ$ around the $\hat{c}$ axis ($\theta=90^\circ$). Solid curve shows a result of curve fit. (b) Fast Fourier-transform (FFT) of the data shown in (a). (c) FFT of the time spectra with $\vec{B}\parallel \hat{c}$ at 5 K. }
\label{fig2}
\end{figure}

Within the $T$ range of the present measurements (2 K $\le T\le$ 290 K), a single diamagnetic state ($\mu^+=$ Mu$^+$) was observed above $\sim$10 K as a signal precessing with a frequency $\nu_0=\gamma_\mu B/2\pi$ (comprising the central line), where  $\gamma_\mu$ ($= 2\pi\times135.53$ MHz/T) is the muon gyromagnetic ratio and $B=|\vec{B}|$. The depolarization rate for this signal was almost independent of $T$ with a Gaussian damping ($\sim$0.02 $\mu$s$^{-1}$); this was unambiguously attributed to random local fields exerted by the nuclear magnetic moments of the $^{47}$Ti (natural abundance: 7.4\%) and $^{49}$Ti (5.4\%) nuclei. Meanwhile, the \msr\ time spectra exhibited a remarkable change below $\sim$10 K. A typical example (observed at 5 K, $B\simeq32.5$ mT) is shown in Fig.~\ref{fig2}(a), where a  beat pattern due to multiple frequency components is clearly observed in the signal amplitude.  

As is evident in Fig.~\ref{fig2}(b),  one can trace two pairs of satellite lines in the fast Fourier-transformed spectra ($\nu_{i\pm}$, $i=1$, 2), which are situated almost symmetrically around the central line ($\nu_0\simeq4.41$ MHz). A preliminary analysis showed that the least-square curve fits of the \msr\ time spectra, assuming one pair of satellites, did not reproduce the data for those obtained with $\vec{B}$ oriented away from the $\langle100\rangle$ direction, yielding a poor $\chi^2$. However, fits with two sets of satellites were found to yield a satisfactory result with significantly improved $\chi^2$ in the majority of cases. It was also found that the splitting ($\nu_{i+}-\nu_{i-}$) of these satellites remained unchanged when the applied field was reduced to 10 mT, indicating that $B$ was in the high field-limit range, where the splitting was independent of the $B$ value.  These observations clearly indicate that a significant fraction of implanted muons form a paramagnetic Mu state characterized by an extremely small HF parameter (almost 10$^{-4}$ times smaller than the vacuum value of $4.463\times10^3$ MHz) with significant anisotropy in the electronic structure.  

\begin{figure}[t]
\includegraphics[width=0.8\linewidth]{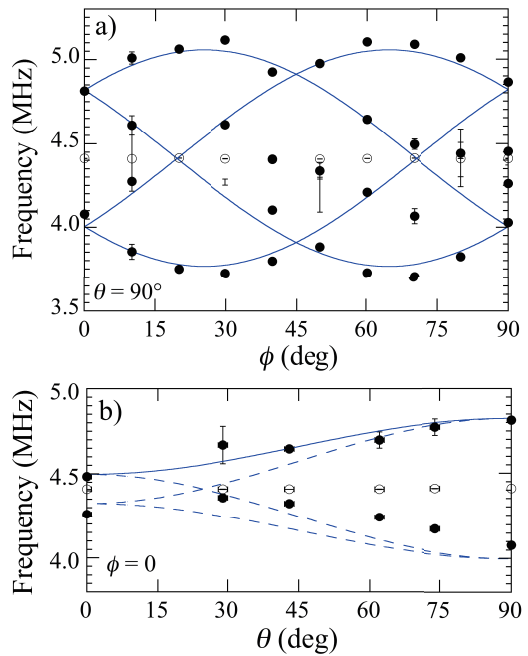}
\caption{(Color online) (a) Angle dependence of muon spin precession frequency observed at 5 K for (a) azimuthal angle ($\phi$) and  (b)  polar angle ($\theta$). Filled symbols: $\nu_{i\pm}$ satellite lines, open symbols: $\nu_0$ central line. The solid curves are obtained via curve fits using general forms of the anisotropic hyperfine parameters, while the dashed curves are calculated for a given set of parameters (see text).}
\label{fig3}
\end{figure}

The results of curve fits for the time spectra assuming multiple precession frequency components ($\nu_{i\pm}$) are summarized in Fig.~\ref{fig3}, for both (a) $\phi$ and (b) $\theta$ dependence. (We assumed that the signal amplitudes of $\nu_{i+}$ and $\nu_{i-}$ were common for each pair in the curve fits.) 
As can be clearly seen in Fig.~\ref{fig3}(a), $\nu_{i\pm}$ exhibits a strong anisotropy with a characteristic angle dependence that is symmetric over $\phi\simeq45^\circ$.  Considering the fact that there are four inequivalent H (Mu) sites with different principal axis orientations for anisotropic HF parameters [see Fig.~\ref{fig1}(b)], four sets of satellite pairs discerned by different $\phi$-dependence can be expected.  
More specifically, it is predicted that the $\phi$ dependence of the HF parameters can be expressed in the form
\begin{eqnarray}
\nu_{i\pm}(\phi)&\simeq&\nu_0\pm\frac{1}{2}A_\perp(\phi),\label{mufrq}\\
A_{\perp}(\phi)&=&A_1\cos^2\Phi+A_2\sin^2\Phi,\label{Aperp}
\end{eqnarray}
where $\Phi=\phi\pm\phi_0$ or $\phi\pm\phi_0\pm\frac{\pi}{2}$, and $\phi_0$ is the offset angle from the $\hat{a}$ axis.
It must be noted that the double sign correspondence between $\nu_{i\pm}$ and the right hand side of Eq.~(\ref{mufrq}) is not unique, because the sign of $A_{\perp}(\phi)$ depends on that of $A_i$ and, thus,  can change over the $\phi$ range.  While our measurements are insensitive to the absolute sign of $A_i$, they can determine the relative sign between $A_1$ and $A_2$. The solid curves shown in Fig.~\ref{fig3}(a) are the result of a simultaneous fit for those curves using common values of $A_i$ and $\phi_0$, where $A_1A_2<0$ is inferred. This strongly suggests that the HF interaction is predominantly determined by the magnetic dipolar field generated by the off-site electron spin(s). (This scenario closely resembles the case of Mu-substituted free radicals often observed in unsaturated organic compounds.)  Moreover, considering that Eqs.~(\ref{mufrq}) and (\ref{Aperp}) predict eight lines when $|A_1|\neq |A_2|$,  Fig.~\ref{fig3}(a) indicates that $|A_1|$ is almost equal to $|A_2|$ within the frequency resolution, which is limited by the observation time range ($\sim$0.1 MHz). The obtained values for $A_i$ and $\phi$ are shown in Table \ref{tab1}. Hereafter, we assume that $A_1<0$ (see below). 

The $\theta$ dependence is shown in Fig.~\ref{fig3}(b), and is characterized by a narrow splitting of $\nu_{1\pm}$ at $\theta=0$ [see Fig.~\ref{fig2}(c)], along with a monotonous increase with increasing $\theta$. Only one pair of satellites has been identified in this data set, most likely because of the narrow splitting relative to the frequency resolution.  It is also notable that $|\nu_{1-}-\nu_0|$ tends to be smaller than $|\nu_{1+}-\nu_0|$ for some unknown reason.  We tentatively attribute this to systematic uncertainty in the present measurement, and adopt $\nu_{1+}$ in order to deduce the HF parameters using a form similar to Eq.~(\ref{Aperp}), such that
\begin{equation}
A_\parallel(\theta) = A_3\cos^2 \theta+A_\perp(0)\sin^2 \theta,
\end{equation}
where the sign of $A_3$ is left undetermined. 
The value of $A_3$ deduced from the curve fit is shown in Table I, where $|A_\perp(0)|$ has been determined  to be 1.314(15) MHz (which is in perfect agreement with the value deduced from the $\phi$ dependence).  

\begin{table}[t]
\begin{tabular}{ccccc}
\hline\hline
  Mu & $A_1$ (MHz) & $A_2$ (MHz) & $A_3$ (MHz) & $\phi_0$ (deg)  \\\hline
$A_{\perp}(\phi)$ & $-1.29(6)$ & $+1.29(6)$ & -- & $25.5(1.4)$ \\
$A_{\parallel}(\theta)$ & -- & -- & $-0.17(2)$  & -- \\\hline\hline
H & -1.276(3)$^*$ & +1.961(3)$^*$ & -1.076(3)$^*$ & 22.1 \\\hline\hline
\end{tabular}
\caption{Mu hyperfine parameters deduced from curve fits of data shown in Fig.~\ref{fig3} (where $|A_1|=|A_2|$ was assumed). The $A_i$'s for the H-related center \cite{Brant:11} are scaled against the gyromagnetic ratio of $\mu^+$ (indicated by an asterisk, see text). The $A_i$ signs for Mu are assumed to be common to those for H. \label{tab1}}
\end{table}

\section{Discussion}
\subsection{Electronic Structure}
It would be intriguing to compare the observed electronic structure with that of a H-related paramagnetic center observed via EPR/ENDOR in a ``lightly reduced" rutile sample \cite{Brant:11}.   According to the proposed model, the center consists of a substitutional Ti$^{3+}$ ion adjacent to a substitutional [OH]$^{-}$ molecular ion (which is identical to the atomic configuration shown in Fig.~\ref{fig1}, apart from the Ti valency).  The HF parameters for the electron-$^1$H nucleus (proton) interaction determined by the ENDOR spectra are reported to be $A_1=-0.401(1)$ MHz, $A_2=+0.616(1)$ MHz, and $A_3=-0.338(1)$ MHz with their principal axes being displaced from the $\langle110\rangle$ or $\langle1\overline{1}0\rangle$ axes by 22.9$^\circ$ (or by 22.1$^\circ$ from the $\hat{a}$ axis). As illustrated in Figs.~\ref{fig1}(c)--(e), the  $A_i$ signs are consistent with the magnetic dipolar field generated by a $d$ electron centered around the Ti site, and their relative magnitudes are in reasonable agreement with those expected for a point-like dipole at the Ti site, i.e., $A_1$:$A_2$:$A_3$ = $-1$:$+2$:$-1$. These values are estimated based on the diagonal terms of the dipole tensor, $A_1\propto-|\overline{\mu}_e|/r^3$, $A_2\propto+2|\overline{\mu}_e|/r^3$, and $A_3\propto-|\overline{\mu}_e|/r^3$ for an effective magnetic moment $|\overline{\mu}_e|$, with $r$ being the distance between the H and Ti atoms. 

Provided that Mu forms a complex state identical to that of H, the corresponding HF parameters can be predicted by simply scaling these values for H using the factor $\gamma_{\mu}/\gamma_{\rm p}=3.1832$ (as shown in Table I, disregarding a minor difference of $\phi_0$), where $\gamma_{\rm p}=2\pi\times42.5774$ MHz/T is the proton gyromagnetic ratio.  Adopting the reasonable assumption that the $A_i$ signs are common between the Mu and H complexes, comparison of the $A_i$ values given in Table I immediately leads us to the conclusion that the Mu complex state differs from that of H. 

In the classical limit, the unpaired electron is regarded as a point-like magnetic dipole situated at the nearest neighboring Ti site, and the HF parameters (per unit Bohr magneton, $\mu_B$) are given by the equation
\begin{equation}
A_i=\frac{\gamma_\mu}{2\pi}g_e\mu_B\frac{3\cos^2\tau_i-1}{2r^3},
\end{equation}
where $g_e$ is the electron $g$ factor, ${\bf r}$ ($r=|{\bf r}|$) is the coordinate of the Ti$^{3+}$ atom with a muon at the origin, and $\tau_i$ is the angle between ${\bf r}$ and the symmetry axis ($\tau_1=\tau_3=\pi/2$, $\tau_2=0$). Assuming that the distance to the nearest neighboring Ti$^{3+}$ atom $r_{\rm nn}=0.226$ nm (i.e., an unrelaxed crystal lattice) and that a full $1\mu_B$ moment (with $g_e=2$) is present, we have $A_1^{\rm nn}=A_3^{\rm nn}=-10.88$ MHz and $A_2^{\rm nn}=+21.77$ MHz, where nn indicates the nearest neighbor.  These values are far greater than the experimental values, which strongly suggests that the spatial distribution must be considered for the spin polarization, e.g., using
\begin{equation}
A_i=\frac{\gamma_\mu}{2\pi}g_e\mu_B\int d^3{\bf r}\rho_{\rm s}({\bf r})\frac{3\cos^2\tau_i-1}{2r^3},\label{SD}
\end{equation}
where  ${\bf r}$ is now the electron coordinate,  and $\rho_{\rm s}$ is the spin density corresponding to the difference between the charge densities of the spin-up and spin-down electron(s) \cite{deWalle:93}.
Comparison of the $A_i^{\rm nn}$ [obtained using Eq.~(\ref{SD}) for $\rho_{\rm s}({\bf r})=\frac{1}{4\pi}\delta(r-r_{\rm nn})$] with the experimental values implies that $\rho_{\rm s}({\bf r})$ may have a considerable spread to reduce the effective moment size $|\mu_e|\simeq\mu_B\int d^3{\bf r}\rho_{\rm s}({\bf r})/r^3$.  

\begin{figure}[t]
\includegraphics[width=0.95\linewidth]{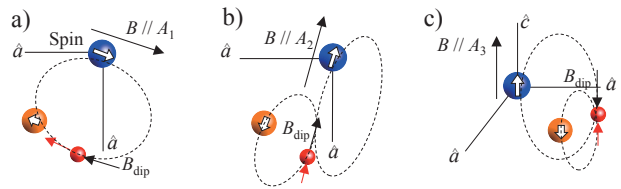}
\caption{(Color online) Hyperfine interactions expected for the magnetic dipolar field ($B_{\rm dip}$) generated by the spin polarization of both the Ti and O atoms under an external field ($\vec{B}$) parallel to the (a) $A_1$, (b) $A_2$, and (c) $A_3$ axes, respectively (see text).}
\label{fig4}
\end{figure} 

Here, we show that both magnitudes and signs of the HF parameters can be understood by considering additional spin polarization of the O atom next to the muon. 
As illustrated in Fig.~\ref{fig4}, assuming that spin polarization at the Ti site induces small polarization of the ligand O antiparallel to the Ti polarization (usually expected for the Heisenberg-type exchange interaction), the $A_i$ incorporating the contribution of the magnetic dipole at the O site are given approximately by 
\begin{eqnarray}
A_1&\simeq&\frac{\gamma_\mu}{2\pi}|\overline{\mu}_e|\left[-\frac{1}{r^3_{\rm nn}}-\frac{2\epsilon}{r^3_{\rm O}}\right],\nonumber\\
A_2&\simeq&\frac{\gamma_\mu}{2\pi}|\overline{\mu}_e|\left[+\frac{2}{r^3_{\rm nn}}+\frac{\epsilon}{r^3_{\rm O}}\right],\label{dipo}\\
A_3&\simeq&\frac{\gamma_\mu}{2\pi}|\overline{\mu}_e|\left[-\frac{1}{r^3_{\rm nn}}+\frac{\epsilon}{r^3_{\rm O}}\right],\nonumber
\end{eqnarray}
where $\epsilon$ is the relative magnitude of the O spin polarization against that of Ti and $r_{\rm O}$ is the OMu bond length (= 0.109 nm).  As summarized in Table \ref{tab2}, simulated  $A_i$ for $|\overline{\mu}_e|=0.050\mu_B$ and $\epsilon=0.077$ (i.e., $\epsilon|\overline{\mu}_e|=0.0039\mu_B$) exhibit almost perfect agreement with the experimental values.  The model also provides a natural explanation for the small magnitude of $A_3$ that the dipole fields from Ti and O atoms nearly cancel at the muon site [see Fig.~\ref{fig4}(c)], suggesting the possibility that the current assumption on the $A_3$ sign may be irrelevant. We note that such residual spin polarization of ligand O atoms is usually expected in the case of strong $d$-$p$ hybridization, as has been observed for a variety of transition metal oxides \cite{Chung:03}.

\begin{table}[t]
\begin{tabular}{cccccc}
\hline\hline
 & $A_1$ (MHz) & $A_2$ (MHz) & $A_3$ (MHz) & $|\overline{\mu}_e|$ &$\epsilon$  \\\hline
Exp. & $-1.29(6)$ & $+1.29(6)$ & $-0.17(2)$ & -- & -- \\
Sim. & $-1.29$ & $+1.09$ & $-0.17$ & $0.050\mu_B$ & 0.077 \\\hline\hline
\end{tabular}
\caption{Comparison of experimental and simulated Mu hyperfine parameters assuming antiferromagnetic spin polarization at nn O site, where the experimental values of $A_1$ and $A_3$ were used to evaluate $A_2$, $|\overline{\mu}_e|$, and $\epsilon$ in the simulation. \label{tab2}}
\end{table}

We currently speculate that the difference in the electronic structure between Mu and H is primarily due to the local environment of the specimen. While the present Mu complex state was observed in an ``as received" crystal, the H-related paramagnetic state was observed after a reduction process (annealing at 600$^\circ$ C for 10 min in a N$_2$ atmosphere \cite{Brant:11}). Considering the fact that the reduction process also yields O vacancies, it may be of interest to recall the recent investigation based on density-functional theory that suggested that H is more stable at the O vacancy site, where it adopts a negatively charged state to form a Ti$^{3+}$H$_{\rm VO}^-$ complex state \cite{Filippone:09}. Thus, it is natural to consider the possibility that the paramagnetic state observed via EPR/ENDOR might correspond to such a state, although the consistency with experimental results must be carefully examined.  The fact that the relative magnitudes of the $A_i$'s for the H-related center correspond to the case of $\epsilon\simeq0$ in Eq.~(\ref{dipo}) (i.e., $A_1$:$A_2$:$A_3$ = $-1$:$+2$:$-1$) seems to constitute supporting evidence, as it suggests that the relevant H atom has no neighboring O atom.  

The fact that the hyperfine interaction is dominated by a magnetic dipolar interaction also places a strong constraint on the diffusion of $\mu^+$ in the $c$ channel, because the $A_i$ are strongly dependent on the distance to the electron ($\propto1/r^3$). The diffusive motion of $\mu^+$ against a stationary $d$ electron at the Ti site would immediately lead to the strong damping of satellite signals due to the fluctuation of $r$ and hence of $A_i$. 

\subsection{Thermal Property}
\begin{figure}[tb]
\includegraphics[height=0.6\linewidth]{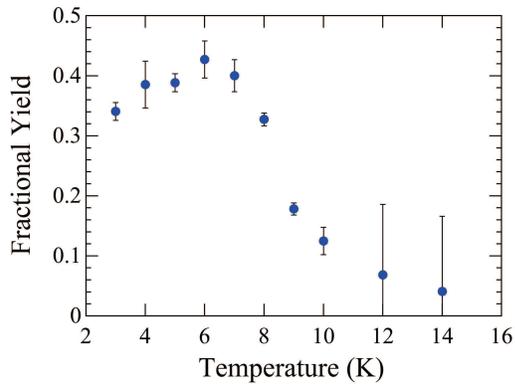}
\caption{(Color online) Temperature dependence of fractional yield for muonium complex state deduced from signal amplitude.}
\label{fig5}
\end{figure}

The $T$ dependence of the fractional yield ($f$) for the Mu complex state is shown in Fig.~\ref{fig5}. The sum of the signal amplitude between the Mu complex state and the diamagnetic state (not shown) is almost independent of $T$, suggesting that the Mu complex is converted to a diamagnetic state above $\sim$10 K.  Here, it must be noted that $f\simeq 0.4$ ($<1$) does not necessarily indicate two different muon sites with different charge states.  In non-metallic compounds, it is usually expected that the initial yield of the paramagnetic state upon muon implantation is predominantly determined by the density of the short-lived free electrons that are produced via radiolysis near the end of the muon radiation track. While this density varies between compounds, it is independent of $T$ because the radiolysis is an athermal process. We assume that the muon site in \tio\ is unique (as shown in Fig.~\ref{fig1}b) over the entire observed $T$ range, where the initial yield of the Mu complex is controlled by the athermal electron density, which is independent of $T$.

In general, the diamagnetic state can be either positively or negatively charged. If the electronic energy levels  ($E_\mu^{0/+}$, $E_\mu^{-/0}$) associated with the Mu complex are situated near the top of the valence band,  a Mu$^-$ state can be expected at lower $T$ because the Fermi level ($E_F$) is likely to be situated far above the mid-gap in the present $n$-type specimen (i.e., $E_\mu^{0/+}<E_\mu^{-/0}<E_F$).  In that case, an {\sl increase} in the yield of the Mu complex would be observed with elevating temperature due to the hole capture process (${\rm Mu}^- +h^+\rightarrow{\rm Mu}$); however, this is contrary to the actual observed behavior.   Thus, the behavior of the Mu complex strongly suggests that the process relevant to the promotion of the diamagnetic state is electron release, ${\rm Mu}\rightarrow{\rm Mu}^+ + e^-$ (i.e., $E_\mu^{0/+}<E_F<E_\mu^{-/0}$), and that the Mu complex state can serve as an electron donor.  

Provided that the ionization of the Mu complex state is driven by an Arrhenius-type activation process,  the disappearance of the Mu signal above $\sim$10 K suggests that the activation energy ($E_a$) is of the order of $10^1$ meV.  This is in line with certain earlier reports suggesting unidentified shallow level states (e.g, via optical absorption \cite{Pascual:78}  or infrared absorption spectroscopy on deuterated rutile\cite{Herklotz:11}). In any case, it must be remembered that the interpretation of $E_a$ depends on the detailed neutral charge state formation process of the Mu complex, and also on the kind of equilibrium state realized for the Mu complex formation.  At one end of the range, $E_a$ represents a direct transition from the defect level to the bottom of the conduction band, while at the other extreme, in equilibrium, it indicates a transition from the defect level to the Fermi level (i.e., $E_a\simeq E_F-E_\mu^{0/+}$). Since the origin of the $n$-type conductivity in the present specimen is unknown, the present value of $E_a$ should be interpreted as a lower bound for the actual defect level.  

\section{Summary and Conclusion}
We have demonstrated that the electronic structure of the interstitial Mu center in rutile is characterized by extremely small and highly anisotropic hyperfine parameters. These parameters are predominantly determined by magnetic dipolar interaction with the unpaired Mu electron, which is primarily situated at the Ti site.  The hyperfine parameters are quantitatively explained by a model that considers a small residual spin polarization of the O atom (which is antiparallel to that of the Ti atom), suggesting that the electronic structure should be interpreted as being a Ti-O-Mu complex state.  The extremely small effective moment size of the unpaired electron ($\sim$0.05$\mu_B$ at the Ti site, $\sim$0.0039$\mu_B$ at the O site) as well as the small activation energy required for its promotion to the conduction band, implies that the Mu complex (and hence the corresponding H state) can serve as an electron donor. This strongly suggests that H is one of the primary origins of unintentional $n$-type conductivity in rutile \tio.

\begin{acknowledgments}
We would like to thank the KEK-MSL staff for their technical support during the muon experiment. We would also like to express our gratitude to K. Yoshizawa, Y. Iwazaki, and S. Tsuneyuki for helpful discussion and for providing us with the result of their first-principle calculation for Mu/H in \tio\ prior to publication. This work was partially supported by the KEK-MSL Inter-University Research Program (2012A0110, 2012B0032).
\end{acknowledgments}

\end{document}